\documentclass{article}
\usepackage[preprint]{spconf}
\copyrightnotice{\copyright\ IEEE 2024}
\toappear{To appear in {\it Proc.\ IEEE ISBI,
   May 27-30, 2024, Athens, Greece.}}

\usepackage{spconf,amsmath,graphicx}
\usepackage{enumitem}
\setlist{nosep, leftmargin=14pt}

\usepackage{mwe} 
\usepackage{tabularx,ragged2e} 
\usepackage{multirow} 
\usepackage{mathtools} 
\usepackage{cite} 

\title{Semi-supervised variational autoencoder for cell feature extraction in multiplexed immunofluorescence images}
\name{
\begin{tabular}{@{}c@{}}
Piumi Sandarenu$^{1}$, 
Julia Chen$^{2,3,4}$,
Iveta Slapetova$^{3}$,
Lois Browne$^{3}$,
Peter H. Graham$^{3,4}$,\\
Alexander Swarbrick$^{2,5}$,
Ewan K.A. Millar$^{4,6,7,8}$,
Yang Song$^{1}$,
Erik Meijering$^{1}$
\end{tabular}
}
\address{$^{1}$ School of Computer Science and Engineering, University of New South Wales, Sydney, Australia\\
$^{2}$ Cancer Ecosystems Program, Garvan Institute of Medical Research, Darlinghurst, Australia.\\
$^{3}$ Cancer Care Centre, St George Hospital, Kogarah, Australia.\\
$^{4}$ St. George \& Sutherland Clinical School, UNSW Sydney, Kensington, Australia.\\
$^{5}$ School of Clinical Medicine, Faculty of Medicine, UNSW Sydney, Australia.\\
$^{6}$ Dept. of Anatomical Pathology, NSW Health Pathology, St. George Hospital, Kogarah, Australia.\\
$^{7}$ Faculty of Medicine \& Health Sciences, Sydney Western University, Campbelltown, Australia.\\
$^{8}$ University of Technology Sydney, Ultimo, Australia.\\
\vspace*{3mm}
}
\begin{document}
%
\maketitle
\begin{abstract}
Advancements in digital imaging technologies have sparked increased interest in using multiplexed immunofluorescence (mIF) images to visualise and identify the interactions between specific immunophenotypes with the tumour microenvironment at the cellular level. Current state-of-the-art multiplexed immunofluorescence image analysis pipelines depend on cell feature representations characterised by morphological and stain intensity-based metrics generated using simple statistical and machine learning-based tools. However, these methods are not capable of generating complex representations of cells. We propose a deep learning-based cell feature extraction model using a variational autoencoder with supervision using a latent subspace to extract cell features in mIF images. We perform cell phenotype classification using a cohort of more than 44,000 multiplexed immunofluorescence cell image patches extracted across 1,093 tissue microarray cores of breast cancer patients, to demonstrate the success of our model against current and alternative methods.
\end{abstract}
\section{Introduction}
\label{sec:intro}


Rapid developments in digital imaging technologies are contributing to growing interest in multiplexed immunofluorescence (mIF) imaging as a tool with potential to revolutionise diagnostics, drug development, and personalised medicine. This technique uses fluorescently labelled antibodies or markers that bind to specific target molecules in formalin fixed paraffin embedded (FFPE) tissue sections allowing in-situ visualisation of multiple immunophenotypes within the tumour microenvironment (TME), leading to accurate assessment of complex biological processes at single-cell resolution. However, mIF images present unique challenges such as spectral overlap, noisy backgrounds, extreme stain variation, complexity, and cost associated with the analysis of spatially resolved high-resolution images with multiple stains. 

Currently, mIF-based TME analysis \cite{Wang2021, KimJunbum2022, VirathamPulsawatdi2020} involves several steps: spectral unmixing, tissue and cell segmentation \cite{QuPathBankhead2017,Cellpose1,Cellpose2}, computation of morphological characteristics (such as shape and size measurements) and biomarker expression statistics in individual cells, nuclei and cytoplasm (using medical image analysis tools), cell clustering, and patient-level analysis carried out using machine learning or statistical approaches \cite{Phenograph, FlowSOM}. 

Due to the noisy background and high stain variability in mIF images, simple segmentation methods based on image processing and basic machine learning tools are not reliable descriptors of nuclei and cell boundaries. Since the cell features used in current methods \cite{Wang2021, KimJunbum2022, VirathamPulsawatdi2020} are directly calculated using these segmentation masks it is possible that these features are not able to accurately represent the complex morphological characteristics and biomarker expression patterns in mIF images. However, training advanced segmentation models require a large amount of manual annotations, incurring substantial resource costs while containing inter- and intra-observer variability. Therefore, we believe that extraction of more comprehensive cell features without requiring precise segmentation can have a direct impact on the subsequent TME analysis stage. Inspired by the success of neural networks, we propose a novel method for cell feature extraction in mIF images that employs a variational autoencoder (VAE) \cite{KingmaVAE2014} with supervision via latent subspace representation. 

Neural networks are widely used for extraction of high-dimensional representations from medical images \cite{TellezNIC2021,LiDSMIL2021,ZhangDTFDMIL2022,ElleryColorectal2020}. An autoencoder is a neural network that maps input images to a lower-dimensional representation using an encoder, often employed as a feature extraction model. VAEs are an improvement on the traditional autoencoder by compressing the latent representations into a probabilistic distribution, thereby forcing the model to learn complex and continuous variations within images. Moreover, supervised autoencoders \cite{LeiLe2018} have been recently introduced as a network that can produce more generalisable representations by using target label prediction as an auxiliary task. However, these models are introduced for generalised computer vision tasks involving natural images which differ significantly from mIF images. 

To achieve a more generalised feature extraction of mIF images, our proposed method uses a VAE as the base model. Furthermore, our model uses \textit{cell phenotype} label as a supervision signal to enhance generalisability and facilitate learning of cell phenotype-related features. However, obtaining cell phenotype labels annotated by expert pathologists is a costly and time consuming task with potentially high inter- and intra-observer variability. In addition, cell phenotype labels can be spuriously correlated with the presence (or absence) of respective biomarkers which can limit the learning of meaningful feature representations due to the inherent inductive biases in neural networks \cite{GeirhosSHORTCUTLEARNING2020, MoayeriMODELSENSITIVITY2022}. To overcome these challenges, we use the labels generated by QuPath \cite{QuPathBankhead2017} and propose to use the full latent representation for image reconstruction and a latent subspace for the joint supervision task. We demonstrate the effectiveness of our approach in generating robust cell feature representations compared to current and alternative methods using a dataset of $n=44,400$ mIF cell images stained using 9 fluorescently labelled antibodies, extracted from 1,093 tissue microarray (TMA) cores belonging to a cohort of 450 breast cancer patients.



\section{Methodology}
\label{sec:method}

\begin{figure}
  \centerline{\includegraphics[width=\columnwidth]{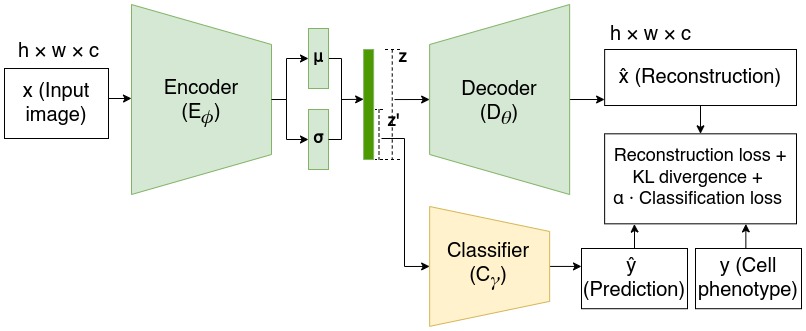}}
  \caption{Proposed VAE framework for cell feature extraction in mIF image data.}
  \label{fig:figSVAE}
\end{figure}

Figure \ref{fig:figSVAE} illustrates the proposed method where we follow the encoder-decoder architecture of a standard VAE \cite{KingmaVAE2014} as our baseline model. Let $X = \{x, y\}$ where $x$ is any image of dimension $h \times w \times c$ and $y$ is the corresponding label. In our experiments, we use mIF image patches of dimension $48 \times 48 \times 9$ as input images and cell phenotype as labels. Let the encoder and decoder networks parameterised by $\phi$ and $\theta$ respectively be denoted as $E_\phi$ and $D_\theta$. Then the encoder network can be written as

\begin{equation}
E_\phi = \biggl(\mu_\phi(x), \text{diag}\Bigl(\sigma_\phi^{2}\bigl((x)\bigr)\Bigr) \biggr)
\label{EQ:encoder}
\end{equation}
which models the distribution $q_\phi(z|x)$ that maps input $x$ to a latent representation $z$. In Eq. (\ref{EQ:encoder}), $\mu_\phi(x)$ and $\text{diag}(\sigma_\phi^{2}((x)))$ represent the latent mean and (diagonal) covariance matrix respectively. The decoder network can be written as
\begin{equation}
D_\theta = \biggl(\mu_\theta(z), \text{diag}\Bigl(\sigma_\theta^{2}(z)\Bigr) \biggr)
\end{equation}
and is used to model the likelihood distribution $p_\theta(x|z)$ which maps the latent representation $z$ back to data space.

We can decompose the joint probability distribution of the VAE $p(x,z)$ into likelihood and prior as $p(x,z)=p(x|z)p(z)$. To infer the true posterior $p(z|x)$, from Bayes theorem we can write $p(z|x)=p(x|z)p(z)/p(x)$ where $p(x) = \int_z p(x,z) dz$. However, calculating $p(x)$ is intractable as it requires integration over all possible configurations of the latent space $z$. Therefore, we maximise the evidence lower bound (ELBO) of the log-likelihood as
\begin{equation} 
\text{ELBO}(\theta, \phi; x) \coloneqq {E}_{z \sim q_{\phi}} \left[\log \frac{p_{\theta}(x,z)}{q_{\phi}(z|x)}\right].
\end{equation}

Since mIF cell images exhibit high variability, we integrate a classifier $C_\gamma$ to increase the generalisability of our model \cite{LeiLe2018}. However, the cell phenotype labels generated using QuPath can be spuriously correlated with the biomarker stains in the mIF cell image which can reduce the complexity of the latent feature space. Therefore, we propose to extract a subspace $z'$ from the latent space $z$ (Fig.~\ref{fig:figSVAE}) where the latent space is now represented as $z=(z - z',z')$. The classifier $C_\gamma$ uses the subspace $\mu_\phi(x)'$ for classification while the full latent space is used for reconstruction. The subspace $\mu_\phi(x)'$ generated by $E_\phi$ can be used to predict the class label probabilities using cross-entropy loss as $p(y|x) = l_\text{CE}\Bigl(y, C_\gamma\left(\mu_\phi(x)'\right)\Bigr)$ and the combined objective function $L(\theta,\phi,\gamma)$ is defined as 
\begin{equation}
L(\theta,\phi,\gamma) \coloneqq - \sum_{(x,y)} \Bigl[ \text{ELBO}(\theta, \phi; x) + \alpha \cdot \log \left(l_\text{CE}\left(y, C_\gamma\right)\right) \Bigr]
\end{equation}
where $\text{ELBO}(\theta, \phi; x)$ is a non-positive quantity, $\alpha$ is a scalar hyperparameter that controls the trade-off between the reconstruction and classification components.

Through this method, we can effectively limit the learning of spurious correlations between labels and input to only a subspace of the full latent representations $z$ and still maintain high reconstruction quality. The size of $z'$ is fixed and dependent on the complexity of the dataset and the level of correlation between the label and input. In our study, we used a proportion of 1/8 for $z'/z$, where $z$ and $z'$ are vectors of size $9,216$ and $1,152$ respectively.

The encoder of our network follows the standard convolutional neural network (CNN) architecture \cite{simonyanVGG} while the decoder consists of transposed convolutional blocks followed by batch normalisation layers and leaky rectified linear unit (ReLU) activation. The loss function of our network incorporates the image reconstruction error (cross entropy loss), Kullback-Leibler (KL) divergence term for regularisation of the latent space \cite{KingmaVAE2014} and classification error (cross entropy loss). The classifier network is comprised of linear layers followed by batch normalisation and leaky ReLU. 

\section{Experiments and results}

\subsection{Dataset and experimental details}
\label{sec:datasetandexp}


Our dataset consists of 18 TMAs with each having 9 channels corresponding to 9 biomarker stains (PD1, CD140b, CD146, Thy1, PanCK, CD8, $\alpha$-SMA, CD31, DAPI) and an additional autofluorescence channel that are all registered. In addition, we also observe that PanCK and CD140b stains are dominant stains in this dataset. Each TMA contains cores of approximately 1.25 \textmu m in diameter scanned at 0.5 \textmu m/pixel resolution. Our dataset consists of 1,093 TMA cores collected from a cohort of 450 breast cancer patients. 

First, we use QuPath's \cite{QuPathBankhead2017} built-in tools based on the watershed algorithm to segment cells and predict cell centres. Following this step, we generate a dataset of $>4.1$ million predictions of cell centres. We depend on image labels generated from QuPath to label the cell phenotypes. Initially, a small number of cell detections are annotated from a single slide for biomarker positivity. Then, we train a classifier using QuPath's built-in tools and use it to classify all cells in the dataset. 

\begin{figure}
\centerline{\includegraphics[width=0.95\columnwidth]{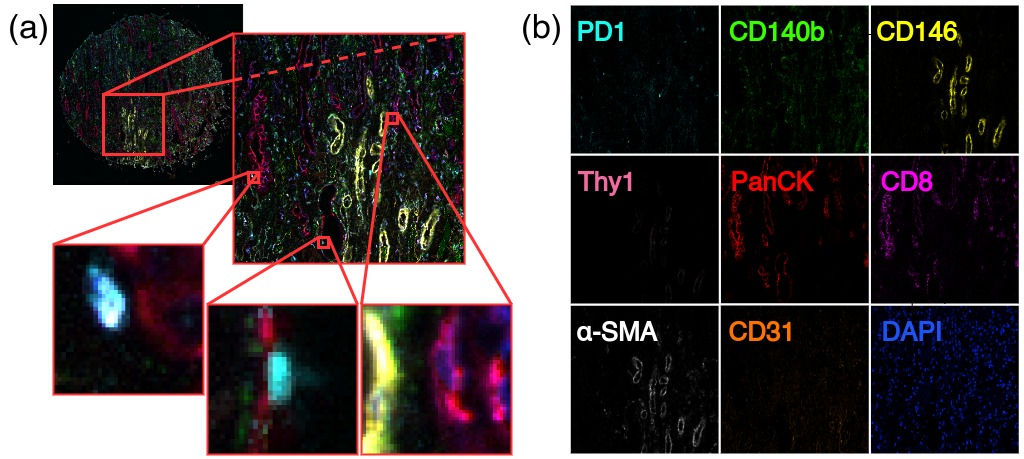}}
  \caption{Example mIF image patches. (a) Expanded views of cells in a mIF TMA core. (b) Each channel of the mIF image captures the presence of a respective biomarker.}
  \label{fig:figPatches}
\end{figure}

To clean the labels, we first remove cells that do not have any biomarker positivity. Then, we remove positive labels for cells whose maximum signal detection (within a cell region segmented using QuPath) is less than 0.5, which can be assumed to be a noisy label. We make further adjustments to narrow down the positive labels for the dominant stains PanCK and CD140b as follows. Since the PanCK stain can be found relatively evenly distributed across the cytoplasm of tumour cells, we remove any cells with positive PanCK detections where the mean expression of PanCK within the cytoplasm region is less than 0.5. Then, we remove positive PanCK and CD140b detections where the maximum cell region is less than 1\% of the maximum of the stain for the respective slide. Finally, the cell phenotypes are categorised based on biomarker positivity predicted by QuPath and the details of this classification and percentage availability is 62.55\% tumour (PanCK+), 22.59\% iCAFs (CD140b+), 7.78\% myCAFs ($\alpha$-SMA+), 3.44\% T-cells (CD8+), 2.50\% dPVLs (CD140+/ CD146+), and 0.92\% exhausted T-cells (PD1+). Due to the significant class imbalance in the dataset, we remove the two classes corresponding to the lowest cell populations in the dataset, which are blood vessels (CD31+) $0.18\%$ and imPVLs (Thy1+) $0.03\%$. 

The final dataset consists of $n=44,400$ randomly selected cell detection representing six cell phenotypes (tumour, iCAFs, myCAFs, T-cells, dPVLs, and exhausted T-cells) in equal proportion across all TMA slides. We use the predicted cell centres as approximations of the actual cell centres and extract cell patches of size $48 \times 48 \times 9$ pixels (Fig.~\ref{fig:figPatches}). We assume that a patch size of $48 \times 48$ pixels, which corresponds to a tissue area of $24 \times 24$ \textmu $\text m^{2}$, is adequate to capture the complete cell image for all cell phenotypes based on the average size of cells in the TME \cite{cellsize2018}. Due to the extreme variability in biomarker intensity within the TMA and across different slides, it is necessary to normalise the images before feeding them as input to our model. We set a lower threshold of 0.5 for the dominant stains PanCK and CD140b while setting a lower threshold of 0.3 for the remaining 7 stains. The upper threshold is set to the maximum intensity observed for each biomarker at slide level. Subsequently, biomarker intensities are scaled using the min-max normalization technique. 

We split the dataset with stratification to allocate $20\%$ as the test set and $80\%$ as the training set, further split as $85\%$ for actual training and $15\%$ for validation. All models are trained with a learning rate of $2\times10^{-5}$ for 1,000 epochs with a batch size of 128 using Tesla V100 GPUs.

\begin{table*}[!t]
\centering
\small
\caption{Comparison of cell classification results for different models.} 
\begin{tabular}{lccccc}
\hline
\bf Method & \bf Image & \bf Embedding & \multicolumn{3}{c}{\bf Test Results (n = 8,880)}\\
\cline{4-6}
& \bf Size & \bf Size & \bf Accuracy & \bf Precision & \bf Recall \\
\hline
ResNet50 Pretrained on ImageNet with PCA & $\ \ 48\times48\times9^{*}$ & 1,152 & 0.7191 & 0.7180 & 0.7215\\
Standard VAE \cite{KingmaVAE2014}& $48\times48\times9$ & 9,216 & 0.8074 & 0.8147 & 0.8125\\
Morphological Features QuPath \cite{QuPathBankhead2017} & - & 156 & 0.8084 & 0.8136 & 0.8088\\
Semi-Supervised Autoencoder \cite{LeiLe2018} & $48\times48\times9$& 9,216 & 0.8201 & 0.8464 & 0.8257\\
Proposed Model & $48\times48\times9$& 1,152 & \bf 0.8486 & \bf 0.8654 & \bf 0.8507\\
\hline
\multicolumn{6}{p{400pt}}{$^*$Features extracted from image patches using pretrained model used as input in the classification task.}
\end{tabular}
\label{resultsGNN}
\end{table*}

\begin{figure}
  \centerline{\includegraphics[width=\columnwidth]{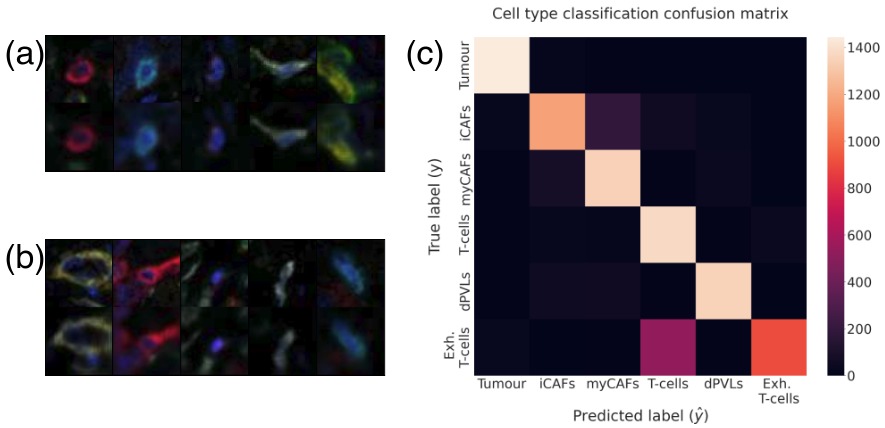}}
  \caption{Reconstruction and classification results. (a,b) Reconstruction results of our model accurately captures cellular features while minimising noise. (c) Confusion matrix shows our model is capable of classifying all cell phenotypes with high accuracy. Exhausted T-cells are more prone to be categorised as T-cells which may be due to the variation in staining strength of CD8+ and PD1+ biomarkers.}
  \label{fig:figCF}
\end{figure}

\begin{table}[ht]
    \centering
    \caption{Comparison of results for the proportion of latent subspace used in the classification network.} 
    \label{resultsproportion}
    \resizebox{1.0\columnwidth}{!}{%
    \begin{tabular}{ccccc}
    \hline
    \bf Proportion & \bf Feature &\multicolumn{3}{c}{ \bf Test Results (n = 8,880) }\\
    \cline{3-5}
    \bf ($z'/z$) & \bf Size ($z'$) & \bf Accuracy & \bf Precision & \bf Recall \\
    \hline
    1  & 9,216 & 0.8377 & 0.8571 & 0.8420 \\
    1/2 & 4,608 & 0.8383 & 0.8645 &  0.8429 \\
    1/8 & 1,152 & \bf 0.8486 & \bf 0.8654 & \bf 0.8507 \\
    1/16 & 576 & 0.8438 & 0.8641 & 0.8462 \\
    \hline
\end{tabular}
}
\end{table}

\subsection{Results}
\label{sec:results}


To evaluate the performance of our method, we compare the cell phenotype classification accuracy using the feature representations of cell patches extracted using different models (Table~\ref{resultsGNN}). Additionally, we present the confusion matrix for 6 class classification as well as some qualitative results in form of cell image reconstructions generated using our model (Fig. \ref{fig:figCF}). For all comparisons we keep the classification network the same except the size of the input layer which corresponds to the size of the feature vector. We train a standard VAE \cite{KingmaVAE2014} and a semi-supervised autoencoder \cite{LeiLe2018} with the same dataset split and use their latent representations for our evaluation. In addition, the results obtained using a ResNet50 pretrained on the ImageNet dataset is presented as a benchmark experiment. We extract a vector of size 2,048 per each channel using the pretrained ResNet50 ($2,048 \times 9$ features per cell patch) and reduce the dimensionality of each channel to 128 using principal component analysis (PCA) ($128 \times 9$ features per cell patch). The features obtained from pretrained ResNet50 with PCA exhibit considerably lower performance, possibly attributing to the substantial dissimilarity in the image domains. To compare the performance of our model to the handcrafted features used in current state-of-the-art methods \cite{Wang2021, KimJunbum2022, VirathamPulsawatdi2020}, we extract 156 important cell features based on the morphological and intensity features of the nuclear segmentation masks. These include 6 contour features for nucleus and cell (such as area, circularity, and eccentricity), nucleus/cell ratio, distance to annotations, and 140 intensity features of each biomarker (such as mean, range, and standard deviation) for nucleus, cell, and cytoplasm regions.

The results of our experiments (Table~\ref{resultsGNN}) confirm that the use of labelled data as a supervisory signal to train a latent subspace of the VAE can be useful in retaining cellular-level features that are relevant to the subsequent tasks in cellular-level analysis of the tumour microenvironment. However, it is important to note that the latent subspace sampling approach is more effective if an optimal subspace size is selected (Table~\ref{resultsproportion}). The selection of this size is subjective of the correlation of the labels and input as well as the quality of the labels available. If the labels are highly correlated with the images, it is recommended to keep the subspace smaller, as less information is required for prediction. We found that the optimal proportion for $z'/z$ in this experiment is 1/8.

\section{Conclusion}
\label{sec:conclusion}

We present a semi-supervised VAE for cell feature extraction in mIF images. We use labels generated by QuPath \cite{QuPathBankhead2017}, in order to reduce the use of exhaustively annotated large datasets of cell images. We propose to use a subsection of the latent space for classification and the full latent space for reconstruction thereby limiting the model from learning spurious correlations between the input and labels. By comparing the results of our method against the current state-of-the-art work, we show that the proposed model is capable of extracting more robust representation of cells. In future work, we aim to expand our experiments to larger datasets and carry out further experiments to evaluate the importance of cellular structures towards patient outcome. We believe our method holds significant potential for advancing research in the analysis of cellular interaction within the TME using mIF images.

\section{Acknowledgments}
\label{sec:acknowledgments}
This research was undertaken with the assistance of resources and services from the National Computational Infrastructure (NCI), supported by the Australian Government.

\section{Ethics approval}
Ethical approval for this study was provided by the South Eastern Sydney Local Health District Human Research Ethics Committee at Prince of Wales Hospital (2018/ETH00138 and HREC 96/16) who granted a waiver of consent to perform research analyses on the tissue blocks. All methods were performed in accordance with the relevant institutional guidelines and regulations.

\bibliographystyle{IEEEtran}
\bibliography{main}

\end{document}